\documentclass[12pt]{article}
\usepackage{psfig,amssymb}
\begin{document}
\title{X-ray Spectral Evolution of Her X-1 in a Low State and the Following Short High State}
\author{S{\i}tk{\i} \,{C}a\u{g}da\c{s} \.{I}nam$^{1}$ \& Altan Baykal$^{2}$ \\ $^{1}$ Department of Electrical and Electronics Engineering,\\ Ba\c{s}kent University, 06530 Ankara, Turkey \\ inam@baskent.edu.tr \\
$^{2}$ Physics Department \\ Middle East Technical University, Ankara 06531, Turkey \\ altan@astroa.physics.metu.edu.tr}

\date{Accepted \ Received}
\maketitle
\begin{abstract}
We analyzed spectral variations of $\sim 8.5$ days long RXTE monitoring observations of Her X-1 in December 2001. This set of observations enables, for the first time, frequent continuous monitoring (111 pointings in $\sim 8.5$ days) of the source with RXTE including $\sim 1.7$ days long low state part and the following $\sim 6.8$ days long short high state part. We used absorbed power law model with iron line energy complex modeled as a Gaussian to fit both the 3-60 keV PCA-HEXTE overall short high state spectrum and 3-20 keV individual PCA spectra. Additional partial cold absorber model was used for both cases. Using 3-20 keV individual PCA spectra, absorption in anomalous dips and preeclipse dips in short high state were compared. Decreasing ratio of unabsorbed flux to absorbed flux with increasing unabsorbed flux in anomalous and preeclipse dips was interpreted as an evidence of the fact that the regions causing opaque obscuration and soft absorption are not geometrically far away from each other. Higher iron line peak energies in low state and short high state ($\sim 6.6-6.9$ keV) were interpreted as a clue of the presence of iron line components other than K$\alpha$ emission line.\\

\end{abstract}
\section{Introduction}
Her X-1, being an eclipsing accretion powered pulsar consisting of a
$\sim 1.24$ s period pulsar in a $\sim 1.7$ day circular orbit, was discovered 
in 1972 with {\it{Uhuru}} observations (Tananbaum et al. 1972). Soon after 
its discovery, its  optical companion was identified to be a blue variable 
thirteenth magnitude star HZ Her (Davidsen al. 1972). The Her X-1/Hz Her 
binary system was shown to display a 35 day long cycle of high and low X-ray flux states. There are two high states, main high and short high, within a single 35 day cycle lasting roughly about 10 and 5 days respectively and separated by $\sim 10$ day long low states (Giacconi et al. 1973; Scott \& Leahy 1999). Superposed on this cycle are eclipses of the neutron star by the companion once per orbital period. While 35 day cycle is mostly coherent since the discovery of the source, there have been four occasions so far when the high states have either failed to turn-on or main high state flux has been reduced (i.e. {\it{anomalous low states}}, see Parmar et al. 1985; Vrtilek et al. 1994; Oosterbroek et al. 2000; Boyd et al. 2004). 

The nature of the 35 day cycle is generally attributed to the periodic variable obscuration of the emission region by a precessing accretion disc viewed nearly edge-on (e.g. Petterson 1975; Scott\& Leahy 1999; Scott, Leahy, Wilson 2000; Leahy 2004). It has been claimed that there is no strict periodicity of this precession (Ogelman 1987): "35 day cycles" generally have durations of 20, 20.5 and 21 orbital cycles with equal statistical probabilities which may lead to the idea that the precession and orbital cycles are related physically (Scott \& Leahy 1999). Baykal et al. (1993) revealed that the statistical interpretation of turn-on behaviour is consistent with a white-noise process in the first derivative of the 35 days phase fluctuations. Other manifestations of 35 day cycle include spectral variations in 35 days (Mihara et al. 1991; Choi et al. 1994B; Leahy 1995; Leahy 2001; Zane et al. 2004), 35 day X-ray pulse profile evolution (Scott et al. 2000), optical pulsations occurring at certain 35 day and orbital phases (Middleditch 1983), and systematic 35 day variations in the optical light curve (Deeter et al. 1976; Gerend \& Boynton 1976).  

Distinct features of high states of Her X-1 in X-ray band are the preeclipse X-ray absorption dips occurring at $\sim 0.8$ orbital phase (Giacconi et al. 1973; Crosa \& Boynton 1980), and anomalous dips (Giacconi et al. 1973; Scott \& Leahy 1999) occurring at $\sim 0.4$ orbital phase. Both types of dips are thought to be due to absorption of cold matter on the line of sight. The occurrence period of the X-ray dips is thought to be related to the beat period of 35 days and 1.7 days ($\sim 1.65$ days), thus they progress to earlier orbital phases within main high and short high states. More detailed features of preeclipse and anomalous dips of Her X-1 were discussed by Leahy et al. (1994), Reynolds \& Parmar (1995), Leahy (1997), Shakura et al. (1998) and Stelzer et al. (1999).

Although considerable changes in the X-ray spectrum in 35 day cycle have been found to be evident, the X-ray spectrum in different 35 day phases can be fitted using a constant-emission model with varying absorption (e.g. Mihara et al. 1991; Leahy 2001). This is another supporting fact that 35 day cycle is due to the precession of an accretion disc that periodically obscures the neutron star beam. In other words, spectral variations in X-ray band are not primarily due to the variations of the primary outgoing beam of the neutron star but rather due to changes in absorption. 

Being the first accretion powered pulsar found to have cyclotron absorption 
feature (Truemper et al. 1978), X-ray spectrum of Her X-1 shows this 
absorption feature around $\sim 40$ keV which indicates a surface magnetic field of $\sim (3.5-4.5)\times 10^{12}$ Gauss. There is no evidence of a second harmonic around $\sim 80$ keV (Coburn et al. 2002).  

Iron line complex at $\sim 6.5$ keV is evident in the X-ray spectrum of Her X-1 (Leahy 2001 and references therein). Furthermore, Endo, Nagase \& Mihara (2000) were able to resolve the feature into two discrete emission lines at $\sim 6.4$ keV and $\sim 6.7$ keV. A distinct Fe XXVI line at $\sim 7$ keV was detected with XMM-Newton throughout 35 day cycle except the main high (Zane et al. 2004). Variations in iron line complex feature over the 35 day period and and the spin period have been evident (e.g. Choi et al. 1994), while variations of this complex feature with the orbital phase have also been studied (e.g. Zane et al. 2004). 

In this paper, results of spectral analysis of December 2001 low state and short high state RXTE data of Her X-1 are presented. This set of observations enables, for the first time, frequent continuous monitoring (111 pointings in $\sim 8.5$ days) of the source with RXTE including $\sim 1.7$ days ($\simeq 1$ orbital period) long low state part and the following $\sim 6.8$ days ($\simeq 4$ orbital periods) long short high state part. Section 2 is an overview about the instruments and the observations. In Section 3, we present the analysis of the observations. In Section 4, results are discussed.

\begin{figure}
\label{fig1_herx1}
\psfig{file=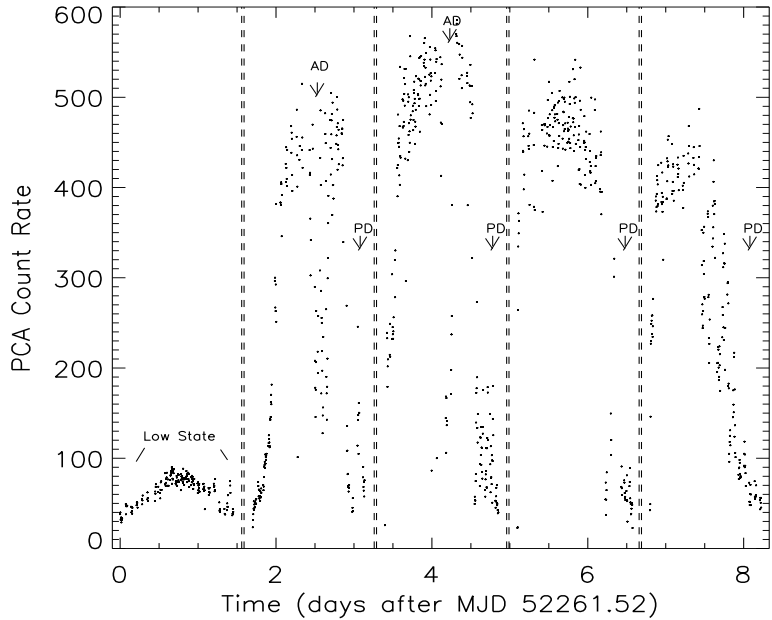,height=9.0cm,width=11.5cm}
\caption{3-20 keV RXTE-PCA Lightcurve of Her X-1. Low state, anomalous dips (AD), and preeclipse dips (PD) are indicated. Vertical dashed lines indicate eclipses of the source.}
\end{figure}

\section{Instruments and Observations}

The RXTE observations used in this paper cover the period from 18 to 26 December 2001 (MJD 52261-52269) with a total exposure of 281.2 ksec (Figure 1). These observations have common proposal ID 60017. The results presented here are based on data collected with the Proportional Counter Array (PCA, Jahoda et al., 1996) and the High Energy X-ray Timing Experiment (HEXTE, Rothschild et al. 1998).
The PCA instrument consists of an array 
of 5 proportional counters (PCU) operating in the 2-60 keV energy range, with 
a total effective area of approximately 6250 cm$^{2}$ and a field of view 
of $\sim 1^{\circ}$ FWHM. Although the number of active PCU's 
varied between 2 and 5 during the observations, observations after 2000 May 13
belong to the observational epoch for which background level
for one of the PCUs (PCU0) increased due to the fact that this PCU started
to operate without a propane layer. The latest combined background models
(CM) are used together with FTOOLS 5.3 to estimate the appropriate PCA background. The HEXTE instrument consists of two independent clusters of detectors, each cluster containing four NaI(TI)/CsI(Na) PHOSWICH scintillation counters (one of the detectors in cluster 2 is not used for spectral information) sharing a common $\sim 1^{\circ}$ FWHM. The field of view of each cluster is switched on and off source to provide background measurements. The net open area of the seven detectors used for spectroscopy is 1400 cm$^2$. Each detector covers the energy range 15-250 keV.

Spectral analysis and error estimation of the spectral parameters are performed 
using {\it{XSPEC}}. The overall 3-60 keV PCA-HEXTE short high spectrum is extracted. In addition to the overall spectrum, individual PCA spectra are extracted in 3-20 keV.  

\section{Spectral Analysis}

\begin{figure}
\label{fig2_herx1}
\psfig{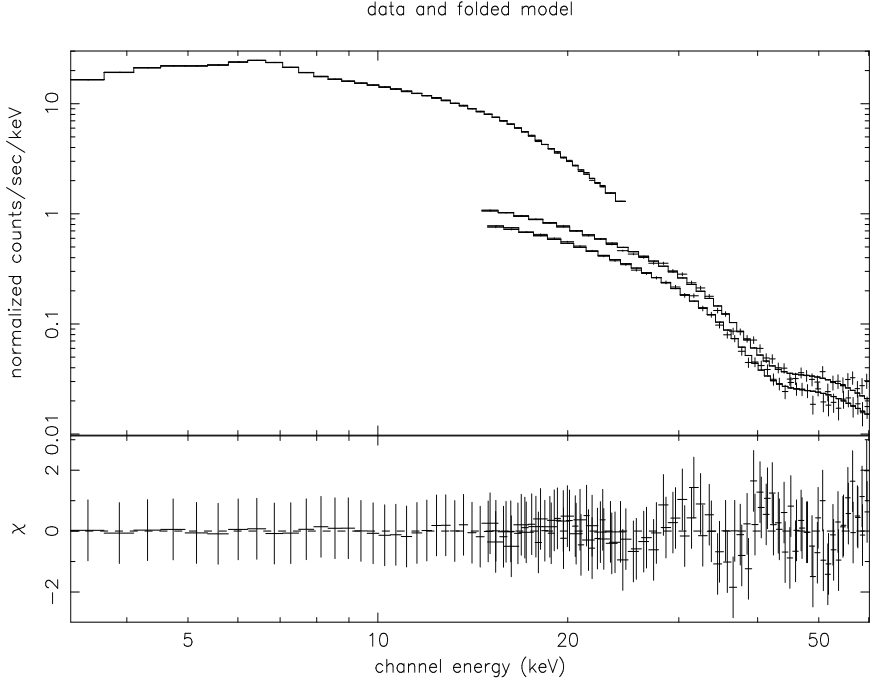}
\caption{Combined PCA and HEXTE spectrum for the short high state of Her X-1. The bottom panel shows the residuals of the fit in terms of $\sigma$ values.}
\end{figure}

2001 December RXTE observation of Her X-1 consists of one low state and four 
short high state orbits with a total exposure of 281.2 ksec. We extract 111 individual PCA spectra in 3-20 keV energy range to investigate the spectral evolution throughout the whole dataset. For the individual PCA spectra, energies lower than 3 keV are ignored due to uncertainties in background modeling while energies higher than 20 keV are ignored as a result of poor counting statistics.  Then, overall 3-60 keV PCA-HEXTE spectrum from the short high state orbits is extracted (Figure 2). We exclude the 3-60 keV spectrum of the low state orbit since the count rate coming from the channels corresponding to $\gtrsim 20$ keV is not statistically significant for both PCA and HEXTE. Additional 2\% systematic error (see Wilms et al. 1999; Coburn et al. 2000) is added to the statistical errors for both the individual spectra and 3-60 keV PCA-HEXTE spectrum.  

\begin{table}
\caption{Sample Spectral Parameters of Individual PCA Observations of Her X-1}
\label{specherx1}  
\[
\tiny{
\begin{tabular}{l c c c c}\hline \hline
Parameter & Low State & Peak of Short High & Anomalous Dip & Preeclipse Dip \\ 
& (t=0.34days) & (t=2.29days) & (t=2.51days) & (t=4.85days) \\ \hline
Hydrogen Column Density & $1.25\mp 0.31$ & $1.54\mp 0.06$ & $1.80\mp 0.34$ & $2.45\mp 0.35$ \\ 
($10^{22}$cm$^{-2}$) & & & & \\
Partial Covering Absorption & & & & \\
-- Hydrogen Column Density & $73.5\mp 11.8$ & $34.6\mp 4.5$ & $69.2\mp 2.8$ &
$121.1\mp 2.9$ \\
($10^{22}$cm$^{-2}$) & & & & \\
-- Partial Cov. Frac. & $0.46\mp 0.07$ & $0.19\mp 0.01$ & $0.83\mp 0.04$ & $0.81\mp
0.04$ \\ 
Iron Line Energy (keV) & $6.72\mp 0.04$ & $6.56\mp 0.02$ & $6.54\mp 0.04$ &
$6.81\mp 0.03$ \\
Iron Line Sigma (keV) & $0.34\mp 0.09$ & $0.53\mp 0.04$ & $1.13\mp 0.05$ & 
$0.74\mp 0.03$\\
Iron Normalization & $0.62\mp 0.10$ & $3.67\mp 0.27$ & $6.74\mp 0.51$ & $1.41\mp
0.11$ \\
(10$^{-3}\times$ cts.cm$^{-2}$.s$^{-1}$) & & & & \\
Power Law Normalization & $0.58\mp 0.02$ & $5.70\mp 0.04$ & $4.11\mp 0.06$ & 
$1.44\mp 0.03$ \\ 
(10$^{-2}\times$ cts.keV$^{-1}$.cm$^{-2}$.s$^{-1}$) & & & & \\ 
3-20 X-ray Flux & & & & \\
(10$^{-10}\times$ ergs.cm$^{-2}$.s$^{-1}$) & & & & \\
--- Absorbed & $1.02\mp 0.07$ & $14.9\mp 0.1$ & $5.54\mp 0.08$ & 
$1.60\mp 0.03$ \\
--- Unabsorbed & $1.37\mp 0.09$ & $16.3\mp 0.1$ & $9.88\mp 0.14$ & $3.37\mp
0.07$ \\
Reduced $\chi^2$ & 1.45 & 0.65 & 0.66 & 1.35 \\ 
(33 d.o.f) & & & & \\ \hline \hline
\end{tabular}}
\]
\vspace{-0cm}
\end{table}

\begin{figure}
\label{fig3_herx1}
\psfig{file=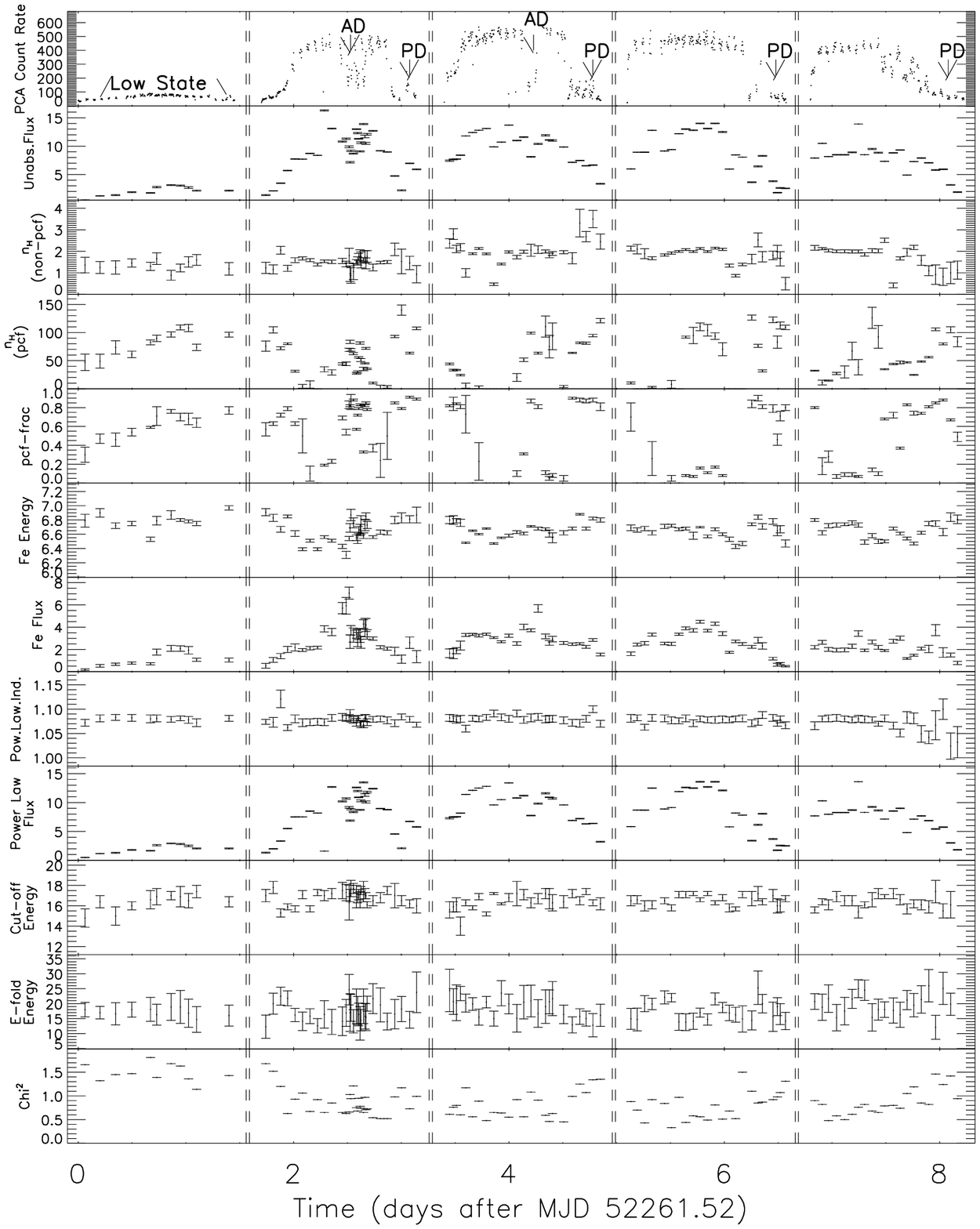,height=20.0cm,width=14cm,angle=0}
\caption{PCA count rate in units of cts/s, 3-20 keV unabsorbed flux in units of 10$^{-10}\times$ ergs.cm$^{-2}$.s$^{-1}$, overall and partial covering hydrogen column densities in units of
10$^{22}$ cm$^2$,  partial covering fraction, iron line energy in units of keV, iron line flux in units of 10$^{-11}\times$ ergs.cm$^{-2}$.s$^{-1}$, power law index, 3-20 keV power law flux in units of 10$^{-10}\times$ ergs.cm$^{-2}$.s$^{-1}$, cut-off and e-folding energies in units of keV, and reduced $\chi^2$ as a function of 
time.  Low state, anomalous dips (AD), and preeclipse dips (PD) are indicated. Vertical dashed lines indicate eclipses. Errors indicate 1$\sigma$ confidence.}
\end{figure}

It is found that the model consisting of an absorbed power law (Morrison \& McCammon, 1983) with an high energy cut-off (White, Swank, Holt 1983) and an iron line feature modeled as a Gaussian is not able to fit the data well. This is expected since two component spectral model with a basic emission spectrum and an absorbed component models or partial covering soft absorption models have been found to fit the main high, short high and low state spectrum of Her X-1 (Mihara et al. 1991; Oosterbroek et al. 1997; Coburn et al. 2000; Oosterbroek et al. 2000; Choi et al. 1994A,B; Leahy 2001). We have found that the model containing partial cold absorber (i.e the model contains both overall absorption represented with the {\it{wabs}} model and partial absorption represented with the {\it{pcfabs}} in XSPEC) fit the 3-20 keV PCA spectra well for both low state and short high state parts of the observation. 

In Table 1, we present sample results of our spectral fits to low state, peak
of short high state, anomalous dip and preeclipse dip data. In Figure 3, we present evolution of spectral parameters during the low state and short
high state orbits.

To fit the overall 3-60 keV PCA-HEXTE short high state spectrum, we used the same spectral model with an additional cyclotron absorption feature at $\sim 42$ keV (Table 2; Figure 2). Her X-1 is the first X-ray pulsar to have shown to have this spectral feature (Truemper et al. 1978). Gruber et al. (2001) showed that the cyclotron resonance energy of Her X-1 changed from $\sim 34$ to $\sim 42$ keV sometime between 1991 and 1993.    

\begin{table}
\caption{Spectral Parameters of PCA-HEXTE Observations of the Short High State of Her X-1}
\label{tabpcahexte}  
\[
\footnotesize{
\begin{tabular}{l c}\hline \hline
Parameter & Value \\ \hline
Hydrogen Column Density ($10^{22}$cm$^{-2}$) & $2.21\mp 0.04$ \\
Partial Covering Absorption & \\
--- Hydrogen Column Density ($10^{22}$cm$^{-2}$) & $65.1\mp 0.8$ \\
--- Partial Covering Fraction & $0.32\mp 0.02$ \\
Gaussian Line Energy (keV) & $6.66\mp 0.02$ \\
Gaussian Line Sigma (keV) & $0.54\mp 0.04$ \\
Gaussian Normalization (cts.cm$^{-2}$.s$^{-1}$) & $(2.35\mp 0.02)\times 10^{-3}$
\\
Power Law Photon Index & $1.09 \mp 0.03$ \\
Cut-off Energy (keV) & $16.5\mp0.2$ \\
E-folding Energy (keV) & $18.7\mp1.5$ \\
Power Law Normalization (cts.keV$^{-1}$.cm$^{-2}$.s$^{-1}$) & $(3.88\mp 0.13)\times 10^{-2}$ \\
Cyclotron Energy (keV) & $42.0\mp 0.2$ \\
Cyclotron Line Width (keV) & $5.13\mp 0.02$ \\
Cyclotron Line Depth & $0.99\mp 0.02$ \\ 
Calculated X-ray Fluxes (3-60 keV in ergs.cm$^{-2}$.s$^{-1}$) & \\
--- Absorbed & $1.22\times 10^{-9}$ \\
--- Unabsorbed & $1.39\times 10^{-9}$ \\
Reduced $\chi^2$ & 1.07 (127 d.o.f) \\ \hline \hline
\end{tabular}}
\]
\vspace{-0cm}
\end{table}

\section{Discussion and Conclusion}

\begin{figure}
\label{fig4_herx1}
\begin{tabular}{ll}
\psfig{file=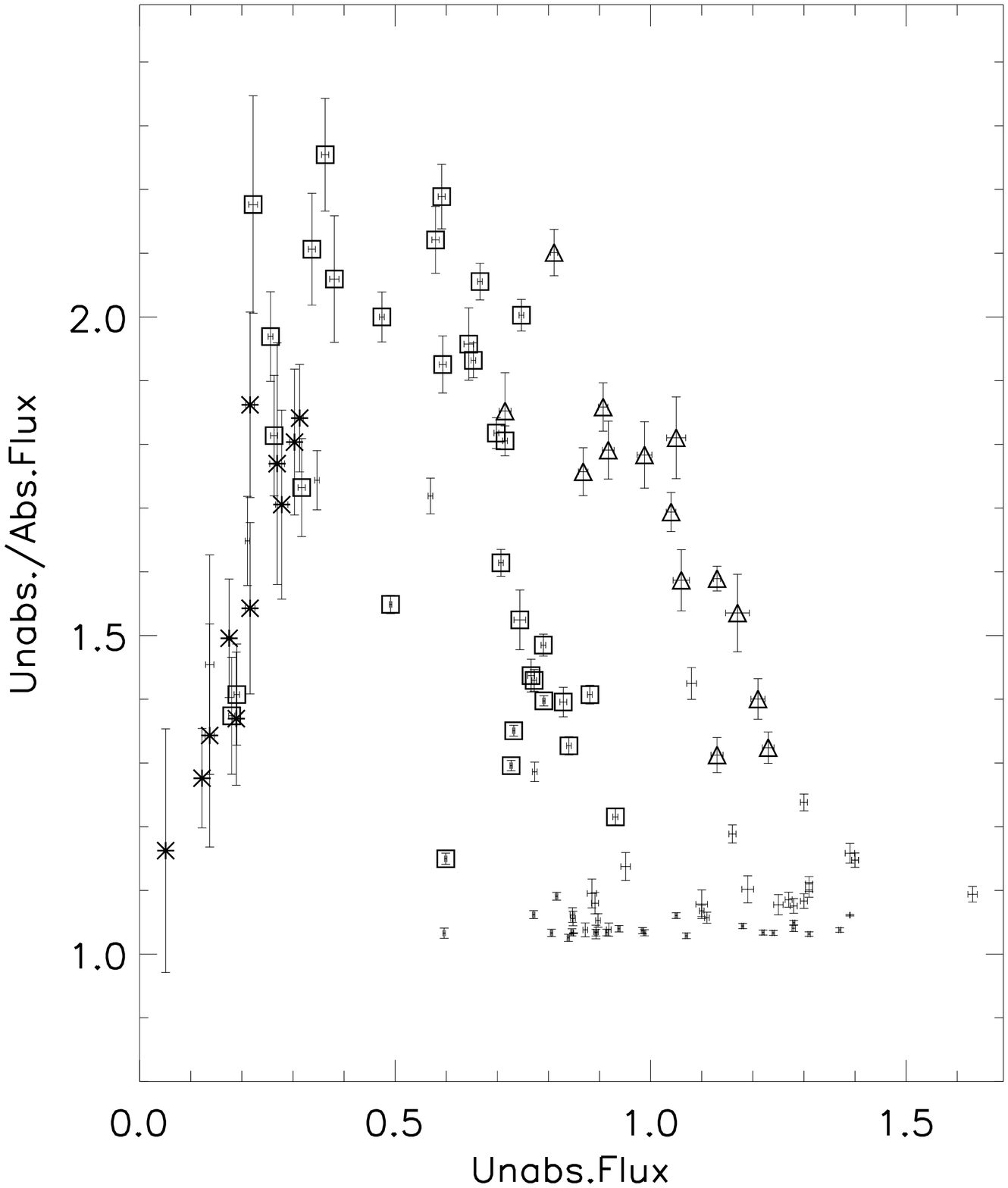,height=7.4cm,width=6.8cm} &
\psfig{file=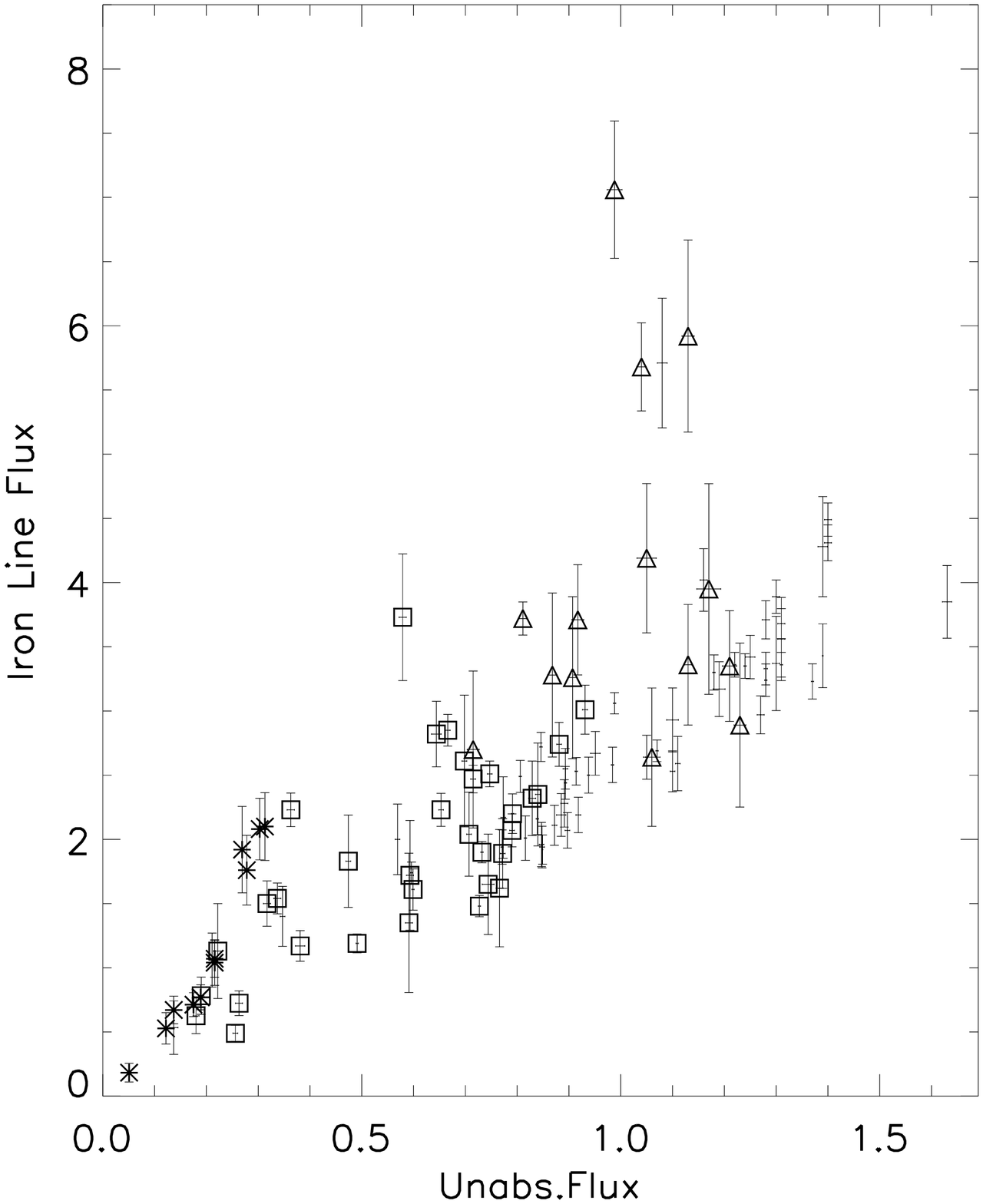,height=7.4cm,width=6.8cm} 
\end{tabular} \\
\caption{Variation of ratio of 3-20 keV unabsorbed flux to 3-20 keV absorbed flux ({\bf{left}}), and variation of iron line flux in units of 10$^{-11}\times$ ergs.cm$^{-2}$.s$^{-1}$ ({\bf{right}}) as a function of 3-20 keV unabsorbed flux in units of 10$^{-10}\times$ ergs.cm$^{-2}$.s$^{-1}$. Asterisks, triangles and squares denote low state, anomalous dip and preeclipse dip, respectively.}
\end{figure}

We analyzed spectra obtained from continuous RXTE monitoring of $\sim 8.5$ 
days of Her X-1. This rather frequent continuous monitoring, for the first time, enabled the study of the spectral evolution of low state, short high peaks and 
short high dips within a relatively compact observation period. 

For the short high state part, stability of the power law index, cut-off energy and e-folding energy throughout the whole monitoring may be an indication of the fact that 
accretion geometry and mass accretion rate does not change considerably. This idea is in accord with the fact that the X-ray flux changes in Her X-1 should primarily be due to the opaque obscuration of the pulsar beam, while there should not be no significant variation in the flux emerging initially from the magnetic poles of the pulsar.

We have found some evidences of differences between anomalous dips and preeclipse dips in short high state orbits. First marginal evidence is the higher ratio of unabsorbed flux to absorbed flux for more than $\sim 50$\% percent of preeclipse dips compared to anomalous dips (left panel of Figure 4). For the rest of the preeclipse dip regions, this ratio is similar to that of anomalous dips. We consider this ratio as a measure of sum of absorptions coming from the total absorption and partial covering absorption (i.e. greater value of this ratio means more absorption). Then, this marginal evidence shows that absorption is more dominant in preeclipse dips compared to anomalous dips. This result is a confirmation of the previous EXOSAT (Reynolds and Parmar 1995) and GINGA (Leahy 1997) results. From EXOSAT observations, it was found that Hydrogen column density was about a few times greater in the preeclipse dips compared to that in the anomalous dips, and that partial covering fractions were higher in preeclipse dips compared to those in the anomalous dips. Similarly, GINGA observations of the absorption dips revealed that $n_H$ was almost an order higher in the preeclipse dips compared to that in anomalous dips. This difference of absorption in two types of absorption dips was explained as a sign of the fact that the intersection of the outer rim of the accretion disc with the line of sight is responsible for the both types of absorption dips, and different absorptions for both types of dips is only due to the difference in orbital phase for these dips (Crosa and Boynton, 1980; Leahy 1997). Another evidence supporting this interpretation was the lack of 35 day dependence on the variation of absorption in absorption dips, which is in accord with the fact that the outer edge of the accretion disc, being not responsible of 35 day cycle, should not be affected by 35 day cycle of precession (Leahy 1997).         
From the left panel of Figure 4, we have also found that unabsorbed flux is higher in anomalous dips compared to preeclipse dips and that the ratio of unabsorbed flux to absorbed flux decreases with increasing unabsorbed flux independently for both preeclipse dips and anomalous dips. Higher unabsorbed flux in anomalous dips is consistent with  the general orbital phase dependence of X-ray flux. 

On the other hand, decreasing ratio of unabsorbed flux to absorbed flux (i.e decreasing absorption) with increasing unabsorbed flux can be understood by the fact that the X-ray flux variation of Her X-1 is the result of opaque obscuration rather than soft absorption (e.g. Still et al. 2001). Variations of unabsorbed flux should be primarily due to the change in opaque obscuration. Opaque obscuration is result of either obscuration of the accretion disc or the eclipsing stellar companion. This relation, therefore, shows that the regions causing opaque obscuration and soft absorption are not geometrically far away from each other.      

Low states in 35 day cycle of Her X-1 is generally believed to be composed of the phases in which X-rays from the neutron star is obscured (Petterson 1975; Scott\& Leahy 1999; Scott, Leahy, Wilson 2000; Leahy 2004). The resultant X-ray flux for both normal low states and anomalous low states is a combination of scattered and reflected components and has been found to fit well except for normalizations and excess absorption to high state spectrum models (e.g. Coburn et al. 2000). Coburn et al. (2000) also found that absorption can be modeled with a cold partial absorber with $n_H$ $\sim 5\times 10^{23}$ cm$^{-2}$ and partial covering fraction of $\sim 0.7$ for both normal and anomalous low states. Results of our low state fits, giving values $\sim (5 - 10)\times 10^{23}$ cm$^{-2}$ and $\sim 0.3 - 0.8$ for $n_H$ and partial covering fraction respectively, are in agreement with those results.
 
From the right panel of Figure 4, iron line flux is correlated, in general, with the X-ray flux for both short high state and low state observations. However, there is a marginal evidence of higher iron line flux for the anomalous dips. Increase in the strength of the iron line feature in the anomalous dips is a natural consequence of the fact that we have some of the X-ray flux coming from the neutron star passing through
the outer rim of the accretion disc which is possibly
enhancing this iron line feature, so the resultant iron line flux increases in the anomalous dips. Although the origin of the anomalous dips and preeclipse dips are similar, the iron line flux is not significantly higher for the preeclipse dips. This may be related to orbital phase difference between anomalous dips and preeclipse dips. Future observations of preeclipse dips of Her X-1 will be useful to study the variation of iron line feature. 
 
Peak energy of the iron line feature is, in general, at $\sim 6.6$ keV for the short high state. From Figure 3, the peak energy tends to increase at the initial short high turn-on (around $\sim 1.9$ days), anomalous dip of the first short high orbit and preeclipse dips. The iron line feature is also found to peak at higher energies ($\sim 6.8$ keV) in the low state. Iron line energy peaking at higher energies is a sign of the fact that this feature does not consist of only the K$\alpha$ Fe emission line. In fact, it has recently been noted that K$\alpha$ Fe line emission tends to be less powerful in short high and low states compared to the main high state (Zane et al. 2004). Zane et al. (2004), using XMM-Newton observations, also has been able to resolve a secondary iron line feature (Fe XXVI) peaking at $\sim 6.9-7.0$ keV which is present only in low state and short high state. Spectral resolution of RXTE-PCA is not good enough to resolve these lines but the higher peak energies of the resultant iron line feature may be a clue of the presence of iron line components other than K$\alpha$ emission line.  

\noindent{{\bf{Acknowledgments}}}

We acknowledge D.M. Scott for his guidance and we thank him for the useful discussions. S.\,{C}.\.{I}. acknowledges the Integrated Doctorate Programme scholarship from the Scientific and Technical Research Council of Turkey (T\"UB\.{I}TAK).

\end{document}